# Point defects in SiC as a promising basis for single-defect, single-photon spectroscopy with room temperature controllable quantum states

Pavel G. Baranov[1, a], Victor A. Soltamov[1], Alexandra A. Soltamova[1]

Georgy V. Astakhov[2], Vladimir V. Dyakonov[2]

[1]Ioffe Physical-Technical Institute, St. Petersburg, 194021 Russia

[2]Experimental Physics VI, Julius-Maximilians University of Würzburg, 97074 Würzburg, Germany

[a] Pavel.Baranov@mail.ioffe.ru, victor_soltamov@mail.ru, astakhov@physik.uni-wuerzburg.de

**Abstract.** The unique quantum properties of the nitrogen–vacancy (NV) center in diamond have motivated efforts to find defects with similar properties in silicon carbide (SiC), which can extend the functionality of such systems not available to the diamond. Electron paramagnetic resonance (EPR) and optically detected magnetic resonance (ODMR) investigations presented here suggest that silicon vacancy ($V_{Si}$) related point defects in SiC possess properties the similar to those of the NV center in diamond, which in turn make them a promising quantum system for single-defect and single-photon spectroscopy in the infrared region. Depending on the defect type, temperature, SiC polytype, and crystalline position, two opposite schemes have been observed for the optical alignment of the ground state spin sublevels population of the $V_{Si}$-related defects upon irradiation with unpolarized light. Spin ensemble of $V_{Si}$-related defects are shown to be prepared in a coherent superposition of the spin states even at room temperature. Zero-field (ZF) ODMR shows the possibility to manipulate of the ground state spin population by applying radiofrequency field. These altogether make $V_{Si}$-related defects in SiC very favorable candidate for spintronics, quantum information processing, and magnetometry.

**Introduction**

Silicon Carbide (SiC) and diamond are examples of wide band gap semiconductors with chemical, electrical and optical properties that make them very attractive for applications under extreme conditions. Until recently, practical applications of semiconductors have been associated with using of defect ensembles. NV defect in diamond - a nitrogen atom substituted for carbon with an adjacent lattice vacancy - is the only known solid-state system where manipulation of the spin states of a single defect was realized by means of optically detected magnetic resonance (ODMR) owing to its unique optical excitation cycle that leads to the optical alignment of a triplet sublevels of the defect ground state. Such systems are the most prominent objects for applications in new generation of supersensitive magnetometers, biosensors, single photon sources [1-7]. The diamond NV defect is in many ways the ideal qubit but it is currently quite difficult to fabricate devices from diamond. It remains difficult to gate these defects electrically. A search for systems possessing unique quantum properties of the NV defect in diamond that can extend the functionality of such systems seems to be a very promising objective. A search to find defects with even more potential (better than excellent [8-10]) has now been launched. SiC was suggested to be able to open up a whole new world of scientific applications.

A convincing point with SiC is that, similar to diamond, the stable spinless nuclear isotopes guarantee long dephasing times. Unusual polarization properties of various vacancy related defects in SiC (exchange-coupled vacancy pairs P3, P5, P6 and P7) were observed by means of electron paramagnetic resonance (EPR) under optical excitation and reported for the first time in the works of Ref. [11], later in Refs. [12-15]. One of the main questions was to establish whether the observed EPR spectra belong to the ground or to excited state (similar problem existed before also for NV defects in diamond). EPR experiments performed at high frequency and at very low temperatures in darkness excluded the possibility of a thermal or optically excited states and as a result it was proved that the EPR spectra of P3, P5, P6 and P7 defects belong to the ground state for all the defects [16,17].

The recent experiments demonstrated [18-22] that several highly controllable defects exist in SiC, and some of them can be manipulated even at room temperatures (RT).

**Results and discussion**. The optically induced inverse population of the defect spin sublevels in 6H-SiC observed at 77 K in Ref.11, labeled as P3, P5, P6, P7, P10 and P11 were identified as Si-C vacancy pair centers in the triplet (S=1) state with zero-field splitting (ZFS) of 43, 9, 449, 442, 17 and $4.10^{-4}$ cm$^{-1}$, respectively. All the centers except P7 have the axial symmetry along $c$-axis. Models showing a possible configurations of $V_{Si}$ –related defects in 6H-SiC lattice in $(11\bar{2}0)$ plane are presented in Fig. 1. P6 and P7 centers were suggested in Ref. 11 to be Si-C vacancy pair with the shortest internal distance, that is along the Si-C bonds. There are two types of the Si-C bonds in SiC: parallel to the $c$-axis and inclined at an angle of $71^0$ giving rise for two possible modification of the defect – P6 and P7 centers (Fig. 1). The other centers with axial symmetry along $c$-axis and smaller values of ZFS were suggested in Ref. [11] to have large internal distances between Si and C vacancies and possible configurations of $V_{Si}$ –related defects, labeled as [$V_{Si}$] in 6H-SiC lattice in $(11\bar{2}0)$ plane are shown in Fig. 1. The EPR spectra with ZFS of P3 and P5 centers were detected by ODMR by monitoring the zero-phonon lines (ZPL's) at 1.433 (V1), 1.398 (V2), and 1.368 eV (V3) in 6H-SiC [13] (ZFL is also included in the designation of the center). Distance between vacancies increases with decreasing zero-field splitting. This paper presents a review of the recent studies of the defects with optically induced inverse population in two main SiC polytypes: 4H-SiC and 6H-SiC. Vacancies into the crystalline SiC lattice were introduced by neutron irradiation (n-irr) with a dose of $10^{15}$ cm$^{-2}$–$10^{20}$ cm$^{-2}$ or by quenching. The EPR spectra were detected at X- (9.3 GHz) and W- (95 GHz) bands on a continuous wave (cw) and pulse spectrometers. A direct-detection EPR (DD-EPR) technique [18] and optically detected magnetic resonance (ODMR) were also used. In DD-EPR experiments sample was selectively excited at 890 nm with 6 ns flashes (ca. 1.5 mJ per flash) from a parametric oscillator LP603 pumped by a Nd-YAG, which was operated at a repetition rate of 11 Hz and signals induced by optical flash are measured in direct absorption and emission mode in the continuous wave regime.

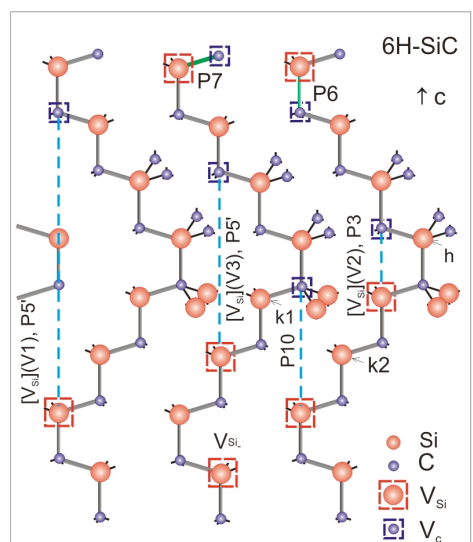

Fig.1 Models showing possible configurations of $V_{Si}$–related defects in 6H-SiC lattice in $(11\bar{2}0)$ plane.

Figure 2 shows EPR spectra of the P6 and P7 centers in n-irr 6H-SiC crystal with dose $10^{20}$ cm$^{-2}$ measured for two orientations (a) by the electron spin echo (ESE) technique at W-band at 1.2 and 1.5 K in darkness, (b) at X-band at 7 K under optical illumination ( n-irr dose of $10^{18}$ cm$^{-2}$). The high-frequency EPR experiments were performed at low temperatures in total darkness, which excluded the possibility of thermal or optical population of the excited state and P6 and P7 centers were concluded to have the triplet ground state. The intensities of the low and high-field fine-structure components measured in the EPR spectra by ESE at temperatures of 1.2 K and 1.5 K sharply differ from each other because of a strong difference in the populations of triplet sublevels at low temperatures and large Zeeman splittings. Intensity ratio of these components gives direct information on the temperature of the sample and allows the positive sign of the fine-structure splitting $D$ to be determined.

Figure 3 shows cw X-band EPR spectra of the $V_{Si}$–related defects detected at 6H-SiC (a) and 4H-SiC (b) with the magnetic field parallel to the *c*-axis. Spectra detected under continuous optical illumination (solid red lines) and without optical illumination (dashed black lines). Vertical bars indicate the positions of the lines for the $V_{Si}$- related defects. The central signal in Figs. 3(a) and 3(b), marked by $V_{Si}^-$, accompanied by a set of hyperfine (HF) lines with a splitting of 0.29 mT and an intensity ratio to the central line of about 0.3. Such signal previously was attributed to the negatively charged silicon vacancy $V_{Si}^-$ with S=3/2 and a fine structure splitting *D* close to zero. From spectra demonstrated on Fig. 3(a) a saturation effect of the signal can be clearly seen in deviation from the first derivative of the EPR signal. Such EPR signals have several contrary explanation. To date three possible models are proposed for such $V_{Si}$ related defect: neutral silicon vacancy $V_{Si}^0$ with S=1, recently revised a low-symmetry modification of the well studied negatively charged silicon vacancy $V_{Si}^-$ in the regular environment with S = 3/2 [14] and previously proposed model where the EPR signals were attributed to the P3 and P5 centers described in [11], which were suggested to be exchanged coupled silicon-carbon vacancy pairs aligned along *c*-axis of the SiC crystal (later P5 signal was shown by ODMR to be associated with two centers which are characterized by almost equal parameters of ZFS). Despite the fact that in [11] P3/P5 were attributed to the triplet center (S= 1) the model can be revised to S=3/2 ground spin state if propose the negatively charged silicon vacancy which interacts with neutral carbon vacancy. Such model can easely describe EPR signals labled as P3, P5, P10, P11 in Ref.11, because the farther away the carbon vacancy located from the $V_{Si}^-$ the smaller *D* splitting to be expected. So the difference of the ZFS values can be explained by different distance between $V_{Si}$ and $V_C$. It is important to underline that the questions of the spin state and hence of the charge state as well as the model of these $V_{Si}$-related defects are still under discussion. But what is seen from Fig. 3 it is that under optical pumping the intensity of the EPR spectra grows substantially and a phase reversal is observed for one of the two transitions in each pair of lines. For instance, $[V_{Si}](V1,V3)$ in 6H-SiC at low temperatures (below 30 K) and $[V_{Si}](V2)$ in 4H-SiC at all temperatures, demonstrate a phase for the high-field transition, while $[V_{Si}](V1,V3)$ in 6H-SiC at high temperatures (above 30 K) and $[V_{Si}](V2)$ in 6H-SiC at all temperatures demonstrate a phase reversal for the low-field transition. As a result of the optical pumping, the distribution of the populations of the spin sublevels in the ground state departs from a Boltzmann distribution. Even an inverse population is created between certain spin sublevels, and emission rather than absorption is detected for one of the transitions (high-field or low-field, depending on the polytype, crystal position, and temperature). Based on a study of the EPR spectra of the $[V_{Si}]$-related defects [16,17] at 95 GHz and low temperatures (1.2–1.5 K) in full darkness, it was unambiguously shown that these spectra belonged to the high-spin ground state and that the sign of the fine-structure splitting *D* is positive. At a temperature of 30 K and with optical excitation the signal $[V_{Si}](V1,V3)$ in 6H-SiC disappears indicating that at this temperature the equal spin sublevel populations are realized.

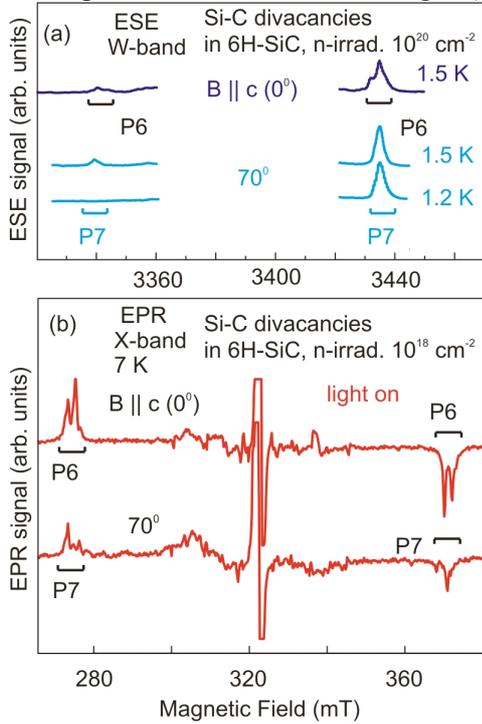

Fig. 2. EPR spectra of the P6 and P7 centers in n-irradiated 6H-SiC crystal measured for two orientations (a) by the ESE technique at W-band at 1.2 and 1.5 K in darkness; (b) at X-band at 7 K under optical illumination.

To explain the photokinetic processes leading to spin alignment under optical pumping, the presence of the excited metastable state is suggested and a spin-dependent intersystem crossing (ICS) between such state, the metastable state and the ground state can be proposed. As was menti-oned above two different models of optical alignment have to be considered for $V_{Si}$-related defects (including divacancies P6 and P7) – with S=1 or *S*=3/2 ground state. The first one is the same as for the NV defect in diamond, because the spin state, as well, as the local symmetry of the defect are similar. The second case with *S*= 3/2 has not been investigated theoretically yet, but some general suggestions can be made with an eye on the triplet model. Shortly $^3$A ($^4$A) ground state, $^3$E ($^4$E) excited state and one $^1$A ($^2$A) metastable state to be needed in order to explain predominant population of *Ms*=0 (*Ms*= ±1/2) spin sublevel, giving rise for the emissive high field transition, if propose that component of the emission decay after optical pumping, related with transition from *Ms*=±1 (*Ms*=±3/2) of $^3$E ($^4$E) to the metastable state $^1$A ($^2$A) is much larger than for the *Ms*=0 (*Ms*=±1/2) of $^3$E ($^4$E) sublevel and further the rates between $^1$A ($^2$A) metastable state and the sublevel $M_S$ =0 (*Ms*=±1/2) of the $^3$A($^4$A) ground state is much larger than the rates of the transitions from $^1$A ($^2$A) to $M_S$ = ±1 (*Ms*= ±3/2) of the ground state. As a result, the spin sublevel $M_S$ =0 (*Ms*=±1/2) of the $^3$A ($^4$A) ground state is predominantly populated, as shown on Fig. 4 (left), and an inverse population is created. This scheme is realized for divacancies P6, P7 and [$V_{Si}$](V1,V3) in 6H-SiC at low temperatures (below 30 K) and for [$V_{Si}$](V2) in 4H-SiC at all temperatures. Another type of optical alignment shown, on Fig. 4 (right), is realized for the $V_{Si}$ –related defect in [$V_{Si}$](V1,V3) and [$V_{Si}$](V2) in the 6H-SiC at temperatures above 30K and as it will be shown further even at RT as well. Here $^3$A ($^4$A) ground state, $^3$E ($^4$E) excited state and metastable $^1$E ($^2$E) state. After optical pumping the rate of the transition from the $M_S$=±1 ($M_S$ =±3/2) of the exited state $^3$E ($^4$E) to the metastable $^1$E ($^2$E) state is again much larger than the rates of the transitions from the $M_S$=0 ($M_S$=±1/2) sublevel, the rates between the state $^1$E ($^2$E) and sublevel $M_S$=0 ($M_S$=±1/2) of the $^3$A ($^4$A) ground state is much less however than the rates of the transitions from $^1$E($^2$E) to the levels $M_S$=±1 ($M_S$=±3/2). As a result, the spin sublevel $M_S$=±1 (±3/2) of the $^3$A ($^4$A) ground state is predominantly populated.

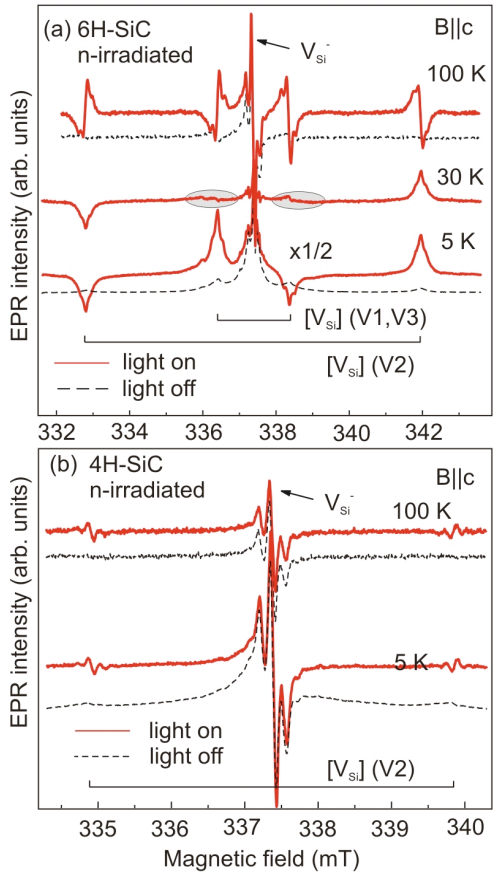

Fig. 3. X-band EPR spectra of the $V_{Si}$- related defects [$V_{Si}$] recorded in 6H-SiC: (a) and in 4H-SiC (b) with the magnetic field parallel to the *c*-axis. Spectra detected under continuous optical illumination – solid red lines, without optical illumination – dashed black lines.

Further investigation of the VSi–related defect properties were produced by means of DD-EPR at temperature T= 300K. Signals of [$V_{Si}$] –related defect recorded at RT in the 4H and 6H-SiC for orientation of the magnetic field perpendicular to the *c*-axis is shown in Fig.5 (top) after excitation of the samples by optical flash into the absorption band of the defect at 890 nm. Under optical pumping, a phase reversal is observed for the [$V_{Si}$](V2) in 4H-SiC even at RT. For 6H-SiC the a phase reversal is observed for [$V_{Si}$](V2) and for [$V_{Si}$](V1,V3) as well. The optical alignment scheme for 4H-SiC is the same as discussed before and shown on Fig.3 (a),(b) though for 6H-SiC we obtain somewhat like RT maser effect as shown on Fig.4.

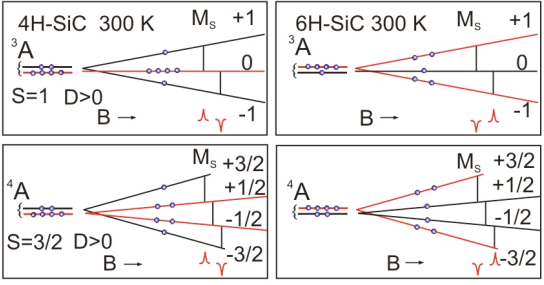

Fig. 4 Two energy-levels schemes for opposite types of optical alignment of the spin sublevels for the ground state of the $V_{Si}$ – related defects in magnetic field on the assumption of the $S=1$ and $S=3/2$ models.

To determine time dependent spin properties of the optically aligned ground spin state of the [$V_{Si}$]–related defect, measurements of the light flash induced DD-EPR signal at the resonant field as a function of time delay after flash were performed. The results of such experiments for the [Vsi](V2) in 4H-SiC are presented on Fig.5(bottom). The transient nutations at RT are shown for three values of microwave power $P$ at resonant magnetic field $B_0$= 321.5mT (low field transition). The transition nutation decays due to inhomogeneity of the B1 microwave field over the sample. In addition, the resonance frequencies are also spread around some mean value of resonance Larmor frequency ‹$\omega_0$› leading to the inhomogeneous line broadening. Clearly, the observed oscillatory behavior demonstrates that the probed spin ensemble can be prepared in a coherent superposition of the spin states at resonant magnetic fields at RT. The population difference of spin states becomes modulated in time with the Rabi frequency given by $\omega_1$= $\gamma B_1$, where $\gamma$ is the gyromagnetic ratio for the electron. The Fourier transforms corresponding to observed oscillations are presented in the insets in Fig.5(bottom). The Rabi frequencies are 0.02, 0.16 and 0.5MHz at P= 30dB, 20dB, and 10 dB, respectively. Rabi oscillations decay with a characteristic time constant $\tau_R$ that depends on the microwave power. Empirically, $\tau_R$ is generally smaller than the spin-spin relaxation time (T2), thus, the lower limit of T2 is about 80 µs at RT.

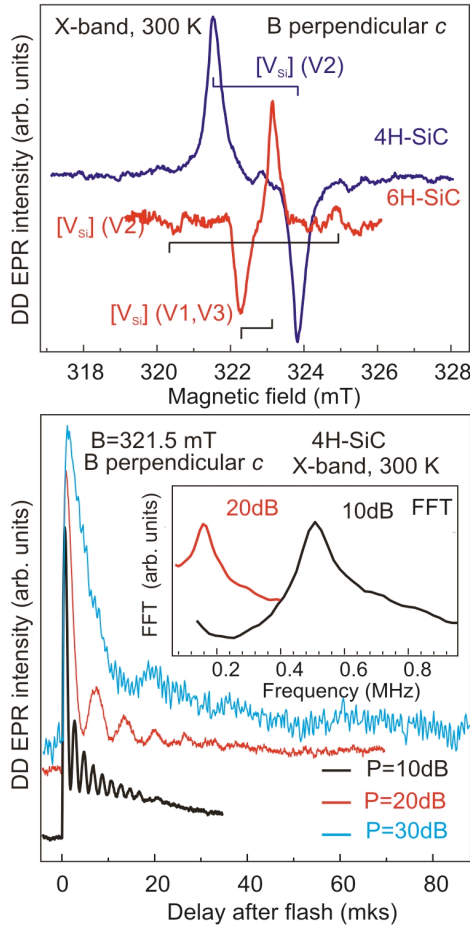

Fig.5 (top) DD-EPR spectra of the $V_{Si}$ –related defects detected in 4H-SiC and 6H-SiC (bottom) at temperature T= 300K. Transient nutations for the defect in 4H-SiC at RT are shown for three values of microwave power. (Inset) Fast Fourier transform (FFT).

In edition low temperature zero-field ODMR experiments on quenched 6H-SiC sample were performed in order to show the possibility of manipulation of the ground state spin population by applying of the radiofrequency which corresponds for the ZFS of the silicon vacancy related defect. Here we demonstrate ZF ODMR effect only for V2 line just as an example. Fluorescence-excitation spectrum of V2 was recorded after single-mode Ti-sapphire laser excitation between 337.0 and 340.17 THz at 4 W/cm$^2$ and shown in Fig.6(a) curve1. Curve 2 in the same figure represents increasing of the fluorescence intensity after resonant radio-frequency field 130MHz have been applied. Spectrum in Fig 6(b) shows the ODMR spectra obtained with excitation of the V2 ZPL and detection at 937 nm. The ODMR spectrum of V2 has its main feature at 130 MHz. As can be seen from the ODMR spectrum resonant radio-frequency field destroyed optically aligned ground state what causes the increasing of the corresponding photoluminescence.

In conclusion, optically induced alignment (polarization) of the ground-state spin sublevels of the $V_{Si}$ – related defect in 4H and 6H-SiC was observed for the first time at RT. In distinction from the known NV defect in diamond, two opposite schemes for the optical spin alignment of $V_{Si}$–related defects in 4H- and 6H-SiC were realized at low temperatures and RT upon illumination with unpolarized light. The alignment schemes depending on the crystal polytype and structure of the $V_{Si}$–related defects which was depicted in Fig. 1. Observed Rabi nutations persist for 80μs at RT and evidence that the probed $V_{Si}$ – related defects spin ensemble can be prepared in a coherent superposition of the spin states at resonant magnetic fields at RT. In addition the electron spin of the defects can be manipulated by low-energy radio field 30-130MHz. Demonstrated spin properties of the defects open up new avenues for quantum computing and magnetometry.

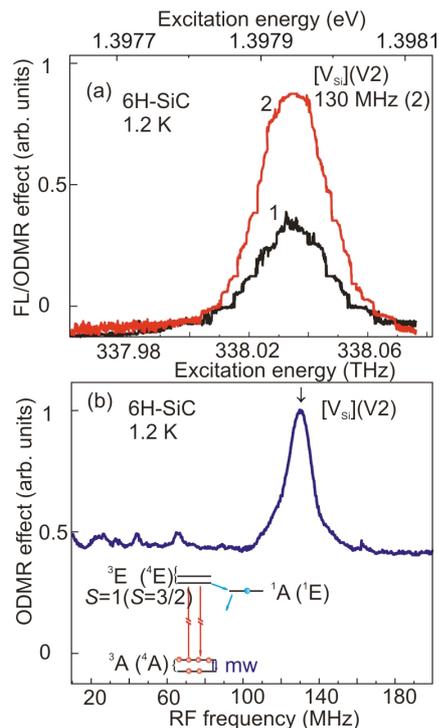

This work was supported by Ministry of Education and Science, Russia, by Contracts the Programs of the Russian Academy of Sciences: "Spin Phenomena in SolidState Nanostructures and Spintronics", "Fundamentals of nanostructure and nanomaterial technologies" № 14.740.11.0048., and by the Russian Foundation for B.R.

Fig.6 (a). Fluorescence-excitation spectra of V2 (1); ODMR excitation spectrum obtained with a resonant rf field at 130 MHz (2);
(b) ODMR spectra obtained with excitation at the *V* 2 ZPL.

**References**

[1] A. Gruber et al., Science 276, 2012–2014 (1997).

[2] F. Jelezko et al., Appl. Phys. Lett. 81, 2160 (2002).

[3] F. Jelezko et al., Phys. Rev. Lett. 92, 076401 (2004).

[4] F. Jelezko and J. Wrachtrup, Phys. Status Solidi A 203, 3207 (2006).

[5]. D.D. Awschalom and M.E. Flatté, Nature Phys. 3, 153–159 (2007).

[6] R. Hanson and D.D. Awschalom, Nature 453, 1043–1049 (2008).

[7] P.M. Koenraad and M.E. Flatté, Nature materials 10, 91 (2011).

[8] J.R. Weber et al., Proc. Natl Acad. Sci. USA 107, 8513 (2010).

[9] D. DiVincenzo, Nature Mat. 9, 468 (2010).

[10] A.G. Smart, Phys. Today 65, 10 (2012).

[11] A.I. Veinger, V.A. Il'in, Yu.M. Tairov, V.F. Tsvetkov, Soviet Physics: Semicond. 13, 1385 (1979); V. S. Vainer and V. A. Il'in, Soviet Physics: Solid State 23, 2126 (1981).

[12] H. J. von Bardeleben et al., Phys. Rev. B 62, 10126 (2000); Phys. Rev. B 62, 10841 (2000).

[13] M. Wagner et al.,, Phys. Rev. B 62, 16555 (2000).

[14] N. Mizuochi et al., Phys. Rev. B 66, 235202 (2002).

[15] W.E. Carlos et al., Phys. Rev. B 74, 235201 (2006).

[16] P.G. Baranov et al., JETP Lett. 82, 441–443 (2005).

[17] S.B. Orlinski et al., Phys. Rev. B 67, 125207 (2003).

[18] P.G. Baranov et al., JETP Letters 86, 202 (2007); Phys. Rev. B 83, 125203 (2011).

[19] W.F. Koehl et al., Nature 479, 84 (2011).

[20] V.A. Soltamov et al., Phys. Rev. Lett. 108, 226402 (2012).

[21] D. Riedel et al., Phys. Rev. Lett. 109, 226402 (2012).

[22] F. Fuchs et al., arXiv:1212.2989 (2012).